\def\rfr#1{eq. (\ref{#1})}

\def\dert#1#2{\frac{{{d}}{#1}}{{{d}}{#2}}}
\def\virg#1{``#1''}

\def\eqi{\begin{equation}}
\def\eqf{\end{equation}}
\def\eqia{\begin{eqnarray}}
\def\eqfa{\end{eqnarray}}
\def\Om{\mathit{\Omega}}
\def\rp#1#2{{#1\over#2}} \def\lb#1{\label{#1}}

\def\kx{\hat{k}_x}
\def\ky{\hat{k}_y}
\def\kz{\hat{k}_z}
\def\bds#1{\boldsymbol{#1}}
\def\co{\cos\omega}
\def\so{\sin\omega}

\def\cO{\cos\Om}
\def\sO{\sin\Om}

\def\cI{\cos I}
\def\sI{\sin I}

\def\ton#1{\left(#1\right)}
\def\qua#1{\left[#1\right]}
\def\grf#1{\left\{#1\right\}}
\def\ang#1{\left\langle #1\right\rangle}

\documentclass[11pt]{article}
\usepackage{url}
\usepackage{amsmath,amsthm,amscd,amssymb}
\usepackage{latexsym}
\usepackage{graphicx,epsfig}
\usepackage{hyperref}
\allowdisplaybreaks[1]

\RequirePackage{color}

\begin{document}

\title{Constraints on Galileon-induced precessions from solar system orbital motions}

\author{L. Iorio \\ Ministero dell'Istruzione, dell'Universit\`{a} e della Ricerca (M.I.U.R.)-Istruzione \\ Fellow of the Royal Astronomical Society (F.R.A.S.) \\
 International Institute for Theoretical Physics and
Advanced Mathematics Einstein-Galilei \\ Permanent address: Viale Unit$\grave{\rm a}$ di Italia 68
70125 Bari (BA), Italy \\ email: lorenzo.iorio@libero.it}

\maketitle

\begin{abstract}
We use latest data from solar system planetary orbital motions to put constraints on some Galileon-induced precessional effects.
Due to the Vainshtein mechanism, the Galileon-type spherically symmetric field of a monopole induces a small, screened correction $\propto \sqrt{r}$ to its usual $r^{-1}$ Newtonian potential which causes a secular precession of the pericenter of a test particle. In the case of our solar system, latest data from Mars allow to constrain the magnitude of such an interaction  down to $\alpha\leq 0.3$ level\textcolor{black}{, where $\alpha$ corresponds to the
non minimal coupling of the Galileon to matter.} Another Galileon-type effect which might impact solar system dynamics is due to an unscreened  constant gradient induced by the peculiar motion of the Galaxy. The magnitude of such an effect, depending on the different gravitational binding energies of the Sun and the planets\textcolor{black}{, taken into account by $\xi$}, is  $\xi \leq 0.004$  from the latest bounds on the supplementary perihelion precession of  Saturn.
\end{abstract}

%\centerline
%{Keywords: Experimental studies of gravity; Experimental tests of gravitational theories;
%Modified theories of gravity; Ephemerides, almanacs, and calendars}

\centerline

\section{Introduction}
Galileon theory\footnote{See, e.g., \cite{rham} for a recent review.} \cite{Nico09} , originally formulated in 4D flat spacetime and later generalized also to curved backgrounds \cite{Def09},  yields a modified  gravity model  relying upon an\footnote{At least, in its original formulation. Multi-Galileon scenarios, introduced for the first time in \cite{Def010}, appeared to cope with certain potential inconsistencies of the single-Galileon model.} additional light scalar field $\pi$, the Galileon, endowed with derivatve self-interactions\footnote{They can help in preserving the observed agreement of the General Theory of Relativity with solar system phenomenology by partly screening the effects of the extra-scalar field at relatively short distances. See below and \cite{CFPS}.}; for a recent overview of this$-$and of many others$-$modified model of gravity, see, e.g., \cite{CFPS}. The name comes from the fact that the Lagrangian for $\pi$ is left unchanged by a transformation of it which generalizes the usual Galilean invariance. The Galileon scenario aims to explain\footnote{See \cite{Fair92a,Fair92b} for an earlier formulation in a different context.} the observed acceleration of the Universe \cite{Riess98,Perlm99,Sper03,Astier06,Wood07,BAO011} without resorting to Dark Energy by modifying the known laws of gravitational interaction at large distances. At the same time, it avoids to compromise their agreement with observations at solar system scales through a screening based on an implementation of the Vainshtein mechanism \cite{Vain}. The Vainshtein mechanism was first shown to work in all its generality \cite{vain1}, and then in the decoupling limit as well \cite{vain2}. Inspired  to the multidimensional braneworld model by Dvali, Gabadadze and Porrati (DGP) \cite{DGP,dgp2,dgp3,dgp4}, the Galileon theory, which has been shown the emerge explicitly from higher-dimensional braneworld scenarios and massive gravity \cite{rham1,rham2,rham3}, is able to cure certain drawbacks of the latter like the appearance of ghosts in the self-accelerated branch \cite{Luty03}. Nonetheless, the single-Galileon scenario is not entirely consistent because of certain issues appearing both at classical and quantum scales, along with unwanted superluminal features \cite{CFPS}. A bi-Galileon theory \cite{Pad011} seems able to successfully face such problems.

In this paper, we want to effectively put quantitative constraints on some consequences of the Galileon scenario with orbital motions in our solar system. \textcolor{black}{For other constraints on Galileon-type interactions inferred from cosmological observations of different phenomena, see, e.g., \cite{ali010,babi011,apple012}.}  As an exciting perspective, local observations can fruitfully shed light on the Dark Energy problem and nature of gravity. Indeed, as we will show in the next Sections, Galileon induces various kinds of local, non-negligible orbital effects, given the present-day level of accuracy in constraining non-standard planetary orbital precessions from solar system observations \cite{fienga011}. In particular, in Section \ref{screnin} we look at the small orbital precessions induced by a single Galileon-type field of a central monopole. In Section \ref{grad} we inspect the effects caused by the unscreened Galileon constant gradient due to the large scale-induced peculiar motion of the Galaxy. As a result, a small differential Sun-planets extra-acceleration should occur in view of their different gravitational self-energies.
For other proposed tests involving different astronomical systems like compact objects in
the Milky Way crossing the Galactic plane, and stars and the central black hole in M87, see \cite{Lam012}. As a general cautionary remark in using certain scenarios less directly accessible and known with respect to our solar system, we want to note that they may typically
suffer from large systematic effects whose accurate
knowledge is often lacking, contrary to Sun's planetary arena. Moreover, in some cases they are also
quite model-dependent in the sense that they heavily rely
upon theoretical assumptions which are still speculative
since they have not yet been tested independently with
a variety of different phenomena, or have not yet been
directly tested at all.
It should also be remarked that some astrophysical
phenomena like the behavior of black holes in extragalctic scenarios  may crucially depend
on the composition, formation and dynamical history
of the systems considered.
This is why we adopt well known and largely
tested orbital motions of some natural bodies in our solar system.
Section \ref{conclu} summarize our findings.
\section{The orbital precessions due to a single Galileon-type field of a central monopole}\lb{screnin}
As a result of the Vainshtein mechanism,  one of the forms in which the Galileon extra-potential for a  localized source of mass $M$  can appear is \cite{Nico09,Burr010,CFPS}
\eqi U_{\rm Gal} = \alpha H_0\sqrt{GM r},\lb{Vpot}\eqf
where $G$ is the Newtonian constant of gravitation and \cite{Hubb011} $H_0=73.8\ {\rm km\ s^{-1}\ Mpc^{-1}}= 2.4\times 10^{-18}\ {\rm s^{-1}}$ is the current value of the Hubble parameter. \textcolor{black}{We emphasize that, in principle, a free parameter-a mass scale-which does not necessarily coincide with $H_0$ can be present in \rfr{Vpot}. Thus, the final
results of our analysis should  depend on such a free parameter and the coupling constant $\alpha$, which is expected to be of the order of unity if the aforementioned free mass parameter does not exactly coincide with $H_0$}.
 \rfr{Vpot} comes from one of the non-linear terms in the spherically symmetric, static equation of motion for $\pi$ near an object of mass $M$ \cite{Nico09,Burr010,CFPS}; such non-linearities become dominant just at relatively short distances. Notably, it is the same non-linear interaction which allows for both the Vainshtein screening and for the self-acceleration of the Universe \cite{Nico09}.

As far as the solar system is concerned, the Galileon becomes dominant with respect to the usual Newtonian potential $U_{\rm N}$ at distances of the order of the Vainshtein radius for the Sun which amounts to about 240 pc. Instead, throughout the planetary region it is
\eqi\left|\rp{U_{\rm Gal}}{U_{\rm N}}\right|=\left|\alpha\right| H_0\sqrt{\rp{r^3}{GM}} \sim 10^{-12}-10^{-9};\lb{scren}\eqf
the screening of the Galileon is, thus, quite effective when the potentials are compared.
Nonetheless, it turns out that \rfr{Vpot} affects planetary orbital motions with secular precessional effects whose expected magnitude may not be hopelessly negligible, being, thus, possible to  effectively  constrain it with present-day observations. Moreover, the expected improvements in the knowledge of the orbital motions of some of the planets of the solar system currently orbited by accurately tracked spacecrafts like MESSENGER (Mercury) and Cassini (Saturn) will further strengthen the bounds on the Galileon-type interaction.

In view of the screening of \rfr{scren}, the Galileon-induced planetary orbital effects can  be  worked out with standard perturbative techniques \cite{BeFa}.

The average of \rfr{Vpot} over a full orbital period $P_{\rm b}$ of a test particle, obtained by using the true anomaly $f$ as fast variable of integration, is
\begin{align}\ang{U_{\rm Gal}} \nonumber & = -\rp{\alpha H_0\sqrt{GM a}}{3\pi}\grf{\sqrt{1-e}\qua{-4E\ton{\rp{2e}{-1+e}} +\ton{1 + e}K\ton{\rp{2e}{-1+e}}} + \right. \\ \nonumber \\
& + \left. \sqrt{1 + e}\qua{-4E\ton{\rp{2e}{1+e}} +\ton{1-e}K\ton{\rp{2e}{1 + e}}}}\lb{Vpotmean},
\end{align}
where\footnote{Here $\pi$ is not the Galileon scalar field, being, instead, the usual ratio of circumference to diameter in Euclidean geometry.} $E$ is the complete elliptic integral and $K$ is the complete elliptic integral of the first kind; $a$ and $e$ are the semimajor axis and the eccentricity of the test particle's orbit.

By using \rfr{Vpotmean} in the Lagrange planetary equations \cite{BeFa} for the variation of the osculating Keplerian orbital elements of a test particle, it follows
\begin{align}
\ang{\dert a t} \lb{dadt} & = 0, \\ \nonumber \\
\ang{\dert e t} \lb{dedt} & = 0, \\ \nonumber \\
\ang{\dert I t} \lb{dIdt} & = 0, \\ \nonumber \\
\ang{\dert \Om t} \lb{dOdt} & = 0, \\ \nonumber \\
\ang{\dert \varpi t} \lb{dpdt} \nonumber & = \rp{\alpha H_0}{2\pi e^2}\grf{\ton{-1 + e}\sqrt{1 + e}\ E\ton{\rp{2e}{e - 1}} -\ton{1+e}\sqrt{1-e}\ E\ton{\rp{2e}{1 + e}} +\right.\\ \nonumber \\
& + \left. \ton{1 - e^2}\qua{\sqrt{1+e}\ K\ton{\rp{2e}{-1+e}} +  \sqrt{1 - e}\ K\ton{\rp{2e}{1 + e}}} }, \\ \nonumber \\
\ang{\dert {\mathcal{M}} t} \lb{dMdt} \nonumber & = \rp{\alpha H_0}{2\pi e^2\sqrt{1-e^2}}\grf{\ton{1+\rp{5}{3}e^2}\qua{\ton{1 - e}\sqrt{1+e}\ E\ton{\rp{2e}{-1+e}}
+\ton{1+e}\sqrt{1-e}\ E\ton{\rp{2e}{1+e}}} -\right.\\ \nonumber \\
& - \left. \ton{1 -\rp{4}{3}e^2 + \rp{e^4}{3} }\qua{\sqrt{1+e}\ K\ \ton{\rp{2e}{-1+e}} +\sqrt{1-e}\ K\ton{\rp{2e}{1+e}}}  },
\end{align}
where $I$ is the inclination of the orbital plane to the reference $\{x,y\}$ plane adopted, $\Om$ is the longitude of the ascending node\footnote{It is an angle in the reference $\{x,y\}$ plane counted from the reference $x$ direction to the line of the nodes, which is the intersection of the orbital plane to the $\{x,y\}$ plane itself.}, $\varpi$ is the longitude of pericenter\footnote{It is a \virg{dogleg} angle since it is defined as $\varpi\doteq\Om +\omega,$ where $\omega$ is the argument of pericenter lying in the orbital plane and counted from the line of the nodes to the point of closest approach.}, and $\mathcal{M}$ is the mean anomaly\footnote{It is defined as $\mathcal{M}\doteq n_{\rm b}\ton{t-t_0},$ where  $n_{\rm b}\doteq \sqrt{GM/a^3}$ is the Keplerian mean motion $t_0$ is the time of passage at pericenter.}.

We remark that \rfr{dadt}-\rfr{dMdt} are exact in the sense that no a-priori simplifying assumptions on the orbital geometry of the test particle were assumed.
Notice also that \rfr{dadt}-\rfr{dMdt} depend on the orbital configuration of the test particle through the eccentricity $e$; they are independent of the semimajor axis $a$.
To order $\mathcal{O}(e^2)$,
\rfr{dpdt}-\rfr{dMdt} reduce to
\begin{align}
\ang{\dert\varpi t} & = -\rp{3\alpha H_0}{8}\ton{1 - \rp{13}{32}e^2}, \\ \nonumber \\
\ang{\dert{\mathcal{M}} t} & =\rp{11\alpha H_0}{8}\ton{1 - \rp{39}{352}e^2}.
\end{align}
We notice that the spherical symmetry of \rfr{Vpot} would  immediately allow to infer that no secular effects at all can occur for $a,e,I,\Om$. Indeed, the Lagrange rate equation for $a$ \cite{BeFa} contains the partial derivative of the averaged perturbing potential $\ang{U_{\rm pert}}$ with respect to $\mathcal{M}$, which is proportional to $t$. The Lagrange equations for $e,I,\Om$ \cite{BeFa} are formed with the partial derivatives of $\ang{U_{\rm pert}}$ with respect to $\Om,\omega, I$ (and $\mathcal{M}$ as well in the Lagrange equation for $e$), which are absent in all spherically symmetric perturbing potentials.

\begin{table*}[ht!]
\caption{Supplementary precessions $\Delta\dot \varpi, \Delta\dot \Om$ of
perihelia  and nodes  of some planets of our solar system %of the solar system
 estimated by Fienga et al. \cite{fienga011} with the INPOP10a ephemerides. Data from Messenger and Cassini were used. All standard Newtonian/Einsteinian dynamics was modeled \cite{fienga011}, with the exception of the solar Lense-Thirring effect. However, it is relevant only for Mercury, given the uncertainties released. The reference $\{x,y\}$ plane is the mean Earth's equator at J$2000.0$. mas cty$^{-1}$ stands for milliarcseconds per century.
}\label{tavolafie}
\centering
\bigskip
\begin{tabular}{lll}
\hline\noalign{\smallskip}
&   $\Delta\dot \Om$ (mas cty$^{-1}$) & $\Delta\dot \varpi $ (mas cty$^{-1}$)  \\
\noalign{\smallskip}\hline\noalign{\smallskip}
Mercury & $1.4 \pm 1.8$ & $0.4 \pm 0.6$ \\
Venus & $0.2 \pm 1.5$ & $ 0.2\pm 1.5$ \\
Earth & $0.0\pm 0.9$ & $-0.2\pm 0.9$ \\
Mars & $-0.05\pm 0.13$ & $-0.04\pm 0.15$ \\
Jupiter & $-40\pm 42$ & $-41\pm 42$ \\
Saturn & $-0.1\pm 0.4$ & $0.15\pm 0.65$ \\
\noalign{\smallskip}\hline\noalign{\smallskip}
\end{tabular}
\end{table*}
From latest Mars data, summarized in Table \ref{tavolafie}, it follows
\eqi\alpha = 0.07\pm 0.3\eqf
\section{Differential orbital precessions induced by unscreened large scale structure Galileon}\lb{grad}
According to Hui and Nicolis \cite{Lam012}, the Galileon symmetry allows for the existence of an additional constant gradient in any solution of the Galileon equation.
In the case of a background external field, such an extra\footnote{In \cite{Lam012}, it is $\pi=\alpha\varphi$.} $\bds\nabla\varphi_{\rm ext}$ would impart an extra-acceleration to a body of mass $M$
\eqi \bds A=-\alpha\ton{\rp{Q}{M}}\bds\nabla\varphi_{\rm ext},\eqf where $Q$ is a conserved scalar charge which coincides with $M$ if the self-gravitational binding energy is negligible; for a compact object like a neutron star $Q/M<1$, while $Q/M=0$ for a black hole.
If, for a given object, such a constant gradient actually exists or not depends on the boundary conditions of the specific scenario considered.
In the case of a galaxy,  they are yielded by the surrounding large scale structure. Hui and Nicolis \cite{Lam012} pointed out that it would be able to induce a long-wavelength scalar field which, on the scale of a galaxy, would resemble just a constant gradient. It would be able to penetrate the Vainshtein zone of the galaxy, acting unsuppressed on it and on all its constituents which, thus, would experience a differential fall according to their  $Q/M$ ratios.

In principle, also a Sun-planet pair should experience such a small differential tug because, although non-relativistic,  our star has a gravitational binding self-energy  much larger than the planets. Thus, it is possible to analyze perturbatively its effects on the orbital motion in detail.
It can be anticipated that, in view of the anisotrpic character of the interaction mediated by $\bds\nabla\varphi_{\rm ext}$, the resulting orbital perturbations should likely  not be limited to just the perihelion and the mean anomaly.
Indeed, the averaged perturbed potential of the relative Sun-planet motion due to the Galileon-induced large scale structure effect is
\begin{align}
\ang{U_{\rm lss}} \lb{Ulss} \nonumber & =\ang{\xi\bds\nabla\varphi_{\rm ext}\bds\cdot\bds r} = -\rp{3}{2}\xi ae A \grf{\co\ton{\kx\cO + \ky\sO} + \right. \\ \nonumber \\
& + \left.\so\qua{\kz\sI +\cI\ton{\ky\cO-\kx\sO}}},
\end{align}
where $\xi$ accounts for the Sun-planet difference in their $Q/M$, and
\begin{align}
A &\doteq \left|\bds\nabla\varphi_{\rm ext}\right|, \\ \nonumber \\
\bds{\hat{k}}&\doteq \rp{\bds\nabla\varphi_{\rm ext}}{\left|\bds\nabla\varphi_{\rm ext}\right|}.
\end{align}
As expected, the Lagrange planetary equations, applied to \rfr{Ulss}, yield the following non-zero secular rates of changes
\begin{align}
\ang{\dert a t}  \lb{dadt2} & = 0, \\ \nonumber \\
\ang{\dert e t} \lb{dedt2}\nonumber & = -\rp{3\xi A\sqrt{1-e^2}}{2n_{\rm b}a}\qua{\kz\sI\co + \cI\co\ton{\ky\cO-\kx\sO} -\right.\\ \nonumber \\
& - \left.\so\ton{\kx\cO +\ky\sO}  }, \\ \nonumber \\
\ang{\dert I t}  \lb{dIdt2}& = \rp{3\xi A e\co\qua{\kz\cI + \sI\ton{\kx\sO-\ky\cO} }}{2n_{\rm b}a}, \\ \nonumber \\
\ang{\dert \Om t} \lb{dOdt2} & = \rp{3\xi A e\csc I\so\qua{\kz\cI + \sI\ton{\kx\sO-\ky\cO}}}{2 n_{\rm b}a\sqrt{1-e^2}}, \\ \nonumber \\
\ang{\dert \varpi t} \lb{dpdt2}\nonumber & = \rp{3\xi A}{2en_{\rm b}a\sqrt{1-e^2}}\grf{\ton{1-e^2}\co\ton{\kx\cO + \ky\sO} +\right.\\ \nonumber \\
& + \left. \so\qua{\ton{\cI-e^2}\ton{\ky\cO-\kx\sO} + \kz\ton{1 - e^2 +\cI }\tan\ton{\rp{I}{2}}   } }, \\ \nonumber \\
\ang{\dert {\mathcal{M}} t} \lb{dMdt2}\nonumber & = -\rp{3\xi A\ton{1 + e^2}}{2en_{\rm b}a}\grf{\co\ton{\kx\cO+\ky\sO} +\right.\\ \nonumber \\
& + \left.\so\qua{\kz\sI +\cI\ton{\ky\cO-\kx\sO}} }.
\end{align}
Also the long-term rates of change of \rfr{dadt2}-\rfr{dMdt2} are exact in the sense that no a-priori simplifying assumptions on the orbital configuration of the test particle were assumed. They are proportional to $\sqrt{a}$, so that they are larger for the outer planets.
For a more realistic contact with actual observations, which are always referred to a specific reference frame, we also did not a-priori align the unit vector $\bds{\hat{k}}$ of the external gradient to any specific direction. It is important to notice that our results \rfr{dadt2}-\rfr{dMdt2} are quite general, so that they do not necessarily refer to a Sun-planet pair, being valid  for other systems including a compact object as well.

The peculiar motion of the Milky Way \cite{Tull08} can be assumed coincident with that of the Local Group with respect to the Cosmic Microwave Background (CMB) rest frame. It occurs at  \cite{Kog93}
\eqi v_{\rm LG} = 627\ {\rm km\ s^{-1}}\lb{veloc}\eqf towards
\begin{align}
{\rm RA} & = 11.11\ {\rm hr}, \\ \nonumber \\
{\rm DEC} & = -27.33\ {\rm deg},
\end{align}
corresponding to
\begin{align}
\kx & = 0.8717,\\ \nonumber\\
\ky & = 0.1711,\\ \nonumber\\
\kz & = -0.459.
\end{align}
The magnitude of the constant gradient can be approximated to the product of the speed of the peculiar motion times the Hubble parameter \cite{Lam012}. In the case of the Milky Way, from \rfr{veloc} we have
\eqi A\sim v_{\rm LG} H_0 = 1.5\times 10^{-12}\ {\rm m\ s^{-2}}.\eqf
From \rfr{dpdt2} and the uncertainty for the supplementary perihelion precession of Saturn in Table \ref{tavolafie} it turns out
\eqi\left|\xi\right| \leq 0.004.\eqf
\section{Summary and conclusions}\lb{conclu}
As far as the Galileon-type spherically symmetric field of a central monopole is concerned, we looked at the  term $\propto H_0\sqrt{r}$ arising from the Vainshtein mechanism. For the Sun and its planets, it is screened to a high level with respect to the usual $r^{-1}$ Newtonian potential; nonetheless, it induces non-vanishing secular precessions of the longitude of perihelion and the mean anomaly which depend only on the planetary eccentricities. Interestingly, their expected magnitudes  are rather close to the present-day level of accuracy in constraining the planetary supplementary precessions, of the order of $\lesssim 1$ milliarcseconds per century. Thus, we were able to effectively constrain the strength parameter $\alpha$ of the residual Galileon interaction down to $\left|\alpha\right| \leq 0.3$ from Mars data.

Then, we looked  at the unscreened Galileon-type local orbital effects caused by the Galactic peculiar motion due to large scale structures. They are proportional to the difference of the gravitational self-energies between the Sun and its planets. Because of the anisotropy of such an interaction, all the osculating Keplerian orbital elements of a test particle undergo non-zero long-term variations, apart from the semimajor axis. By using the supplementary perihelion precession of Saturn, we were able to infer $\left|\xi\right|\leq 0.004$ for the strength parameter of such an interaction.

Further gathering and processing data from ongoing and forthcoming spacecraft-based missions orbiting some planets of the solar system like Mercury (MESSENGER, BepiColombo) and Saturn (Cassini) will allow to enhance such thrilling opportunity of shedding light on crucial cosmological issues from local observations.
\section*{Acknowledgments}
I thank S. Deser and C. de Rham for having pointed out to me relevant references. I am also grateful to an anonymous referee for competent and useful remarks.
\bibliography{SMEbib,Anellobib,Operabib,Vainbib}{}

\providecommand{\href}[2]{#2}\begingroup\raggedright\begin{thebibliography}{10}

\bibitem{rham}
C.~de~Rham, ``Galileons in the sky,'' {\em Comptes Rendus de l'Acad\'{e}mie des
  Sciences} (2012) , \href{http://arxiv.org/abs/1204.5492}{{\ttfamily
  arXiv:1204.5492 [astro-ph.CO]}}. at press.

\bibitem{Nico09}
A.~Nicolis, R.~Rattazzi, and E.~Trincherini, ``Galileon as a local modification
  of gravity,'' {\em Physical Review D} {\bfseries 79} no.~6, (2009) 064036.

\bibitem{Def09}
C.~Deffayet, S.~Deser, and G.~Esposito-Far\`{e}se, ``Generalized galileons: All
  scalar models whose curved background extensions maintain second-order field
  equations and stress tensors,'' {\em Physical Review D} {\bfseries 80} no.~6,
  (2009) 064015.

\bibitem{Def010}
C.~Deffayet, S.~Deser, and G.~Esposito-Far\`{e}se, ``Arbitrary p-form
  galileons,'' {\em Physical Review D} {\bfseries 82} no.~6, (2010) 061501.

\bibitem{CFPS}
T.~Clifton, P.~Ferreiera, A.~Padilla, and C.~Skordis, ``Modified gravity and
  cosmology,'' {\em Physics Reports} {\bfseries 513} no.~1-3, (2012) 1--189.

\bibitem{Fair92a}
D.~B. Fairlie, J.~Govaerts, and M.~A., ``Universal field equations with
  covariant solutions,'' {\em Nuclear Physics B} {\bfseries 373} no.~1, (1992)
  214--232.

\bibitem{Fair92b}
D.~B. Fairlie and J.~Govaerts, ``Euler hierarchies and universal equations,''
  {\em Journal of Mathematical Physics} {\bfseries 33} no.~10, (1992)
  3543--3566.

\bibitem{Riess98}
A.~G. Riess, A.~V. Filippenko, P.~Challis, A.~Clocchiatti, A.~Diercks, P.~M.
  Garnavich, R.~L. Gilliland, C.~J. Hogan, S.~Jha, R.~P. Kirshner,
  B.~Leibundgut, M.~M. Phillips, D.~Reiss, B.~P. Schmidt, R.~A. Schommer, R.~C.
  Smith, J.~Spyromilio, C.~Stubbs, N.~B. Suntzeff, and J.~Tonry,
  ``Observational evidence from supernovae for an accelerating universe and a
  cosmological constant,'' {\em The Astronomical Journal} {\bfseries 116}
  no.~3, (1998) 1009--1038.

\bibitem{Perlm99}
S.~Perlmutter, G.~Aldering, G.~Goldhaber, R.~A. Knop, P.~Nugent, P.~G. Castro,
  S.~Deustua, S.~Fabbro, A.~Goobar, D.~E. Groom, I.~M. Hook, A.~G. Kim, M.~Y.
  Kim, J.~C. Lee, N.~J. Nunes, R.~Pain, C.~R. Pennypacker, R.~Quimby,
  C.~Lidman, R.~S. Ellis, M.~Irwin, R.~G. McMahon, P.~Ruiz-Lapuente, N.~Walton,
  B.~Schaefer, B.~J. Boyle, A.~V. Filippenko, T.~Matheson, A.~S. Fruchter,
  N.~Panagia, H.~J.~M. Newberg, W.~J. Couch, and {The Supernova Cosmology
  Project}, ``Measurements of $\omega$ and $\lambda$ from 42 high-redshift
  supernovae,'' {\em The Astrophysical Journal} {\bfseries 517} no.~2, (1999)
  565--586.

\bibitem{Sper03}
D.~N. Spergel, L.~Verde, H.~V. Peiris, E.~Komatsu, M.~R. Nolta, C.~L. Bennett,
  M.~Halpern, G.~Hinshaw, N.~Jarosik, A.~Kogut, M.~Limon, S.~S. Meyer, L.~Page,
  G.~S. Tucker, J.~L. Weiland, E.~Wollack, and E.~L. Wright, ``First-year
  wilkinson microwave anisotropy probe (wmap) observations: Determination of
  cosmological parameters,'' {\em The Astrophysical Journal Supplement Series}
  {\bfseries 148} no.~1, (2003) 175--194.

\bibitem{Astier06}
P.~Astier, J.~Guy, N.~Regnault, R.~Pain, E.~Aubourg, D.~Balam, S.~Basa, R.~G.
  Carlberg, S.~Fabbro, D.~Fouchez, I.~M. Hook, D.~A. Howell, H.~Lafoux, J.~D.
  Neill, N.~Palanque-Delabrouille, K.~Perrett, C.~J. Pritchet, J.~Rich,
  M.~Sullivan, R.~Taillet, G.~Aldering, P.~Antilogus, V.~Arsenijevic,
  C.~Balland, S.~Baumont, J.~Bronder, H.~Courtois, R.~S. Ellis, M.~Filiol,
  A.~C. Gon\c{c}alves, A.~Goobar, D.~Guide, D.~Hardin, V.~Lusset, C.~Lidman,
  R.~McMahon, M.~Mouchet, A.~Mourao, S.~Perlmutter, P.~Ripoche, C.~Tao, and
  N.~Walton, ``The supernova legacy survey: measurement of $\omega_{\rm m}$,
  $\omega_{\Lambda}$ and $w$ from the first year data set,'' {\em Astronomy and
  Astrophysics} {\bfseries 447} no.~1, (2006) 31--48.

\bibitem{Wood07}
W.~M. Wood-Vasey, G.~Miknaitis, C.~W. Stubbs, S.~Jha, A.~G. Riess, P.~M.
  Garnavich, R.~P. Kirshner, C.~Aguilera, A.~C. Becker, J.~W. Blackman,
  S.~Blondin, P.~Challis, A.~Clocchiatti, A.~Conley, R.~Covarrubias, T.~M.
  Davis, A.~V. Filippenko, R.~J. Foley, A.~Garg, M.~Hicken, K.~Krisciunas,
  B.~Leibundgut, W.~Li, T.~Matheson, A.~Miceli, G.~Narayan, G.~Pignata, J.~L.
  Prieto, A.~Rest, M.~E. Salvo, B.~P. Schmidt, R.~C. Smith, J.~Sollerman,
  J.~Spyromilio, J.~L. Tonry, N.~B. Suntzeff, and A.~Zenteno, ``Observational
  constraints on the nature of dark energy: First cosmological results from the
  essence supernova survey,'' {\em The Astrophysical Journal} {\bfseries 666}
  no.~2, (2007) 694--715.

\bibitem{BAO011}
C.~Blake, T.~Davis, G.~B. Poole, D.~Parkinson, S.~Brough, M.~Colless,
  C.~Contreras, W.~Couch, S.~Croom, M.~J. Drinkwater, K.~Forster, D.~Gilbank,
  M.~Gladders, K.~Glazebrook, B.~Jelliffe, R.~J. Jurek, I.-H. Li, B.~Madore,
  D.~C. Martin, K.~Pimbblet, M.~Pracy, R.~Sharp, E.~Wisnioski, D.~Woods, T.~K.
  Wyder, and H.~K.~C. Yee, ``The wigglez dark energy survey: testing the
  cosmological model with baryon acoustic oscillations at $z= 0.6$,'' {\em
  Monthly Notices of the Royal Astronomical Society} {\bfseries 415} no.~3,
  (2011) 2892--2909.

\bibitem{Vain}
A.~Vainshtein, ``To the problem of nonvanishing gravitation mass,'' {\em
  Physics Letters B} {\bfseries 39} no.~3, (1972) 393--394.

\bibitem{vain1}
C.~Deffayet, G.~Dvali, G.~Gabadadze, and A.~Vainshtein, ``Nonperturbative
  continuity in graviton mass versus perturbative discontinuity,'' {\em
  Physical Review D} {\bfseries 65} no.~4, (2002) 044026.

\bibitem{vain2}
A.~Nicolis and R.~Rattazzi, ``Classical and quantum consistency of the dgp
  model,'' {\em Journal of High Energy Physics} {\bfseries 2004} no.~06, (2004)
  059.

\bibitem{DGP}
G.~Dvali, G.~Gabadadze, and M.~Porrati, ``4d gravity on a brane in 5d minkowski
  space,'' {\em Physics Letters B} {\bfseries 485} no.~1-3, (2000) 208--214.

\bibitem{dgp2}
A.~Lue and G.~Starkman, ``Gravitational leakage into extra dimensions: Probing
  dark energy using local gravity,'' {\em Physical Review D} {\bfseries 67}
  no.~6, (2003) 064002.

\bibitem{dgp3}
A.~Lue, ``The phenomenology of dvali-gabadadze-porrati cosmologies,'' {\em
  Physics Reports} {\bfseries 423} no.~1, (2006) 1--48.

\bibitem{dgp4}
G.~Dvali, A.~Gruzinov, and M.~Zaldarriaga, ``The accelerated universe and the
  moon,'' {\em Physical Review D} {\bfseries 68} no.~2, (2003) 024012.

\bibitem{rham1}
C.~de~Rham and A.~Tolley, ``Dbi and the galileon reunited,'' {\em Journal of
  Cosmology and Astroparticle Physics} {\bfseries 2010} no.~05, (2010) 015.

\bibitem{rham2}
C.~de~Rham and G.~Gabadadze, ``Selftuned massive spin-2,'' {\em Physics Letters
  B} {\bfseries 693} no.~3, (2010) 334--338.

\bibitem{rham3}
C.~de~Rham and G.~Gabadadze, ``Generalization of the fierz-pauli action,'' {\em
  Physical Review D} {\bfseries 82} no.~4, (2010) 044020.

\bibitem{Luty03}
M.~A. Luty, M.~Porrati, and R.~Rattazzi, ``Strong interactions and stability in
  the dgp model,'' {\em Journal of High Energy Physics} {\bfseries 2003}
  no.~09, (2003) 029.

\bibitem{Pad011}
A.~Padilla, P.~M. Saffin, and S.-Y. Zhou, ``Bi-galileon theory ii:
  phenomenology,'' {\em Journal of High Energy Physics} {\bfseries 2011} no.~1,
  (2011) 99.

\bibitem{ali010}
A.~Ali, R.~Gannouji, and M.~Sami, ``Modified gravity \`{a} la galileon: Late
  time cosmic acceleration and observational constraints,'' {\em Physical
  Review D} {\bfseries 82} no.~10, (2010) 103015.

\bibitem{babi011}
E.~Babichev, C.~Deffayet, and G.~Esposito-Far\`{e}se, ``Constraints on
  shift-symmetric scalar-tensor theories with a vainshtein mechanism from
  bounds on the time variation of g,'' {\em Physical Review Letters} {\bfseries
  107} no.~25, (2011) 251102.

\bibitem{apple012}
S.~Appleby and E.~Linder, ``Galileons on trial,''
  \href{http://arxiv.org/abs/1204.4314}{{\ttfamily arXiv:1204.4314
  [astro-ph.CO]}}.

\bibitem{fienga011}
A.~Fienga, J.~Laskar, P.~Kuchynka, H.~Manche, G.~Desvignes, M.~Gastineau,
  I.~Cognard, and G.~Theureau, ``The inpop10a planetary ephemeris and its
  applications in fundamental physics,'' {\em Celestial Mechanics and Dynamical
  Astronomy} {\bfseries 111} no.~3, (2011) 363--385.

\bibitem{Lam012}
L.~Hui and A.~Nicolis, ``An observational test of the vainshtein mechanism,''
  \href{http://arxiv.org/abs/1201.1508}{{\ttfamily arXiv:1201.1508
  [astro-ph.CO]}}.

\bibitem{Burr010}
C.~Burrage and D.~Seery, ``Revisiting fifth forces in the galileon model,''
  {\em Journal of Cosmology and Astroparticle Physics} {\bfseries 2010} no.~08,
  (2010) 011.

\bibitem{Hubb011}
L.~Riess, Adam G. nad~Macri, S.~Casertano, H.~Lampeitl, H.~C. Ferguson, A.~V.
  Filippenko, S.~W. Jha, W.~Li, and R.~Chornock, ``A 3$\%$ solution:
  Determination of the hubble constant with the hubble space telescope and wide
  field camera 3,'' {\em The Astrophysical Journal} {\bfseries 730} no.~2,
  (2011) 119.

\bibitem{BeFa}
B.~{Bertotti}, P.~{Farinella}, and D.~{Vokrouhlick\'{y}}, {\em Physics of the
  Solar System}.
\newblock Kluwer Academic Press, Dordrecht, 2003.

\bibitem{Tull08}
R.~B. Tully, E.~J. Shaya, I.~D. Karachentsev, H.~M. Courtois, D.~D. Kocevski,
  L.~Rizzi, and A.~Peel, ``Our peculiar motion away from the local void,'' {\em
  The Astrophysical Journal} {\bfseries 676} no.~1, (2008) 184--205.

\bibitem{Kog93}
A.~Kogut, C.~Lineweaver, G.~F. Smoot, C.~L. Bennett, A.~Banday, N.~W. Boggess,
  E.~S. Cheng, G.~de~Amici, D.~J. Fixsen, G.~Hinshaw, P.~D. Jackson,
  M.~Janssen, P.~Keegstra, K.~Loewenstein, P.~Lubin, J.~C. Mather, L.~Tenorio,
  R.~Weiss, D.~T. Wilkinson, and E.~L. Wright, ``Dipole anisotropy in the cobe
  differential microwave radiometers first-year sky maps,'' {\em The
  Astrophysical Journal} {\bfseries 419} (1993) 1--6.

\end{thebibliography}\endgroup
%-----------------------------------------

\end{document}